\RequirePackage{tikz}
\documentclass[pdflatex, sn-standardnature]{sn-jnl}% Standard Nature Portfolio Reference Style
\usepackage{amsthm,amsmath}

\usepackage{pdfpages}
\usepackage{enumerate}
\usepackage{tabularx}
\usepackage{lineno}
\usepackage{makecell}
\usepackage{bbm}

\usepackage{amsmath,graphicx,centernot}

\jyear{2024}%

\theoremstyle{thmstyleone}%
\theoremstyle{thmstyletwo}%

\usepackage{bm} % to bold italic mathematical font
\usepackage{breqn} % break lines in equation
\usepackage{systeme}  % to use linear equations
\usepackage{amssymb} 
\DeclareMathOperator{\E}{E}
\DeclareMathOperator{\expit}{expit}
\DeclareMathOperator{\logit}{logit}

\usepackage{multirow}

\theoremstyle{thmstylethree}%

\usetikzlibrary{shapes,decorations,arrows,calc,arrows.meta,fit,positioning}
\tikzset{
    -Latex,auto,node distance =1 cm and 1 cm,semithick,
    state/.style ={ellipse, draw, minimum width = 0.7 cm},
    point/.style = {circle, draw, inner sep=0.04cm,fill,node contents={}},
    bidirected/.style={Latex-Latex,dashed},
    el/.style = {inner sep=2pt, align=left, sloped}
}
\tikzstyle{arrow} = [thick,->,>=stealth]

\begin{document}

\title[QuasiMed]{A Quasi-Regression Method for the Mediation Analysis of Zero-Inflated Single-Cell Data}


\author*[1]{\fnm{Seungjun} \sur{Ahn}}\email{seungjun.ahn@mountsinai.org}
\author[2]{\fnm{Donald} \sur{Porchia}}\email{donald.porchia@ufl.edu}
\author[3,4]{\fnm{Panos} \sur{Roussos}}\email{panagiotis.roussos@mssm.edu}
\author[5]{\fnm{Maaike} \sur{van Gerwen}}\email{maaike.vangerwen@mountsinai.org}
\author[2]{\fnm{Qing} \sur{Lu}}\email{lucienq@ufl.edu}
\author[2]{\fnm{Zhigang} \sur{Li}}\email{zhigang.li@ufl.edu}

\affil[1]{\orgdiv{Department of Population Health Science and Policy}, \orgname{Icahn School of Medicine at Mount Sinai}, \city{New York}, \state{NY}, \country{U.S.A}}
\affil[2]{\orgdiv{Department of Biostatistics}, \orgname{University of Florida}, \city{Gainesville}, \state{FL}, \country{U.S.A}}
\affil[3]{\orgdiv{Department of Psychiatry}, \orgname{Icahn School of Medicine at Mount Sinai}, \city{New York}, \state{NY}, \country{U.S.A}}
\affil[4]{\orgdiv{Department of Genetics and Genomic Sciences}, \orgname{Icahn School of Medicine at Mount Sinai}, \city{New York}, \state{NY}, \country{U.S.A}}
\affil[5]{\orgdiv{Department of Otolaryngology - Head and Neck Surgery}, \orgname{Icahn School of Medicine at Mount Sinai}, \city{New York}, \state{NY}, \country{U.S.A}}


\abstract{\textbf{Motivation:} Recent advances in single-cell technologies have advanced our understanding of gene regulation and cellular heterogeneity at single-cell resolution. Single-cell data contain both gene expression levels and the proportion of expressing cells, which makes them structurally different from bulk data. Currently, methodological work on causal mediation analysis for single-cell data remains limited and often requires specific distributional assumptions.
\\
\textbf{Results:} To address this challenge, we present QuasiMed, a mediation framework specialized for single-cell data. Our proposed method comprises three steps, including (i) screening mediator candidates through penalized regression and marginal models (similar to sure independence screening), (ii) estimation of indirect effects through the average expression and the proportion of expressing cells, (iii) and hypothesis testing with multiplicity control. The key benefit of QuasiMed is that it specifies only the mean functions of the mediation models through a quasi-regression framework, thereby relaxing strict distributional assumptions. The method performance was evaluated through the real-data-inspired simulations, and demonstrated high power, false discovery rate control, and computational efficiency. Lastly, we applied QuasiMed to ROSMAP single-cell data to illustrate its potential to identify mediating causal pathways. \\
\textbf{Availability:} R package is freely available on GitHub repository at \url{https://github.com/sjahnn/QuasiMed}. \\}

%%================================%%
%% Sample for structured abstract %%
%%================================%%




\maketitle

\section{Introduction}\label{smore:introduction}
High-dimensional mediation analysis has been emerging in recent biomedical research, where large numbers of molecular features may act as intermediate variables between an exposure and an outcome. Over the past decade, statistical mediation methods have been extended to handle multiple mediators in high-dimensional settings \citep{intro.mediation1, intro.mediation2}, typically adopting the counterfactual framework for defining direct and indirect effects \citep{intro.counterfactual1}. Recent work has introduced partial sum statistic approaches with sample splitting to facilitate global testing of indirect effects and mediator prioritization in high-dimensional settings \citep{newmethod2025}. Mediation frameworks have also been adapted to a range of omics data types, including microbiome \citep{microbiome1,microbiome2}, epigenetic \citep{epigenetic1, epigenetic2, JS.study1}, metabolomic \citep{metabolomics1}, and transcriptomic studies \citep{transcriptomic1}. Many of these methods rely on parametric models tailored to the distributional features of specific data types. Single-cell RNA sequencing (scRNA-seq, hereafter) data pose additional challenges for mediation analysis. Unlike bulk RNA sequencing, which aggregates gene expression across heterogeneous cell populations, scRNA-seq measures transcript abundance in individual cells and typically yields high-dimensional and sparse data with substantial cell-to-cell variability. In addition to average expression levels \citep{startingref1, startingref2}, the proportion of cells expressing a given gene provides additional biological information on activation and cellular heterogeneity. It provides novel insights into cellular composition and transcriptomes in neurobiology (brain and its function) \citep{scRNAneuro} and many types of cancer \citep{scRNAcancer}.

We recently proposed the first causal mediation framework for scRNA-seq data based on beta and negative-binomial models, called MedZIsc \citep{MedZIsc}. MedZIsc specifies parametric models for both the proportion of cells expressing a gene and its average expression level to account for sparsity and overdispersion. While parametric modeling can be efficient when correctly specified, single-cell data often exhibit gene-specific mean-variance patterns that are difficult to capture with a single parametric family \citep{scrna1}. Moreover, distributional misspecification may affect estimation and inference in high-dimensional settings.

To address these limitations, we develop a quasi-regression-based mediation framework for scRNA-seq data. The method does not assume specific parametric distributions for the mediators and instead models their regression relationships directly. It jointly considers average expression and the proportion of cells expressing a gene, and incorporates a screening step to reduce computational burden in high-dimensional settings. The proposed framework is further distinguished from existing causal mediation methods including the aforementioned methods (e.g., HIMA \cite{JS.study1}), in two important respects. First, while most existing approaches target the natural indirect effect, our method instead targets the interventional indirect effects (IIE) \citep{IIEref}, which relies on less assumptions and is therefore more applicable in practice. Second, our framework is designed to handle zero-inflation, a key data characteristic intrinsic to scRNA-seq data. We refer to the proposed framework as \textbf{QuasiMed}, a quasi-regression approach for mediation analysis in single-cell transcriptomic data. QuasiMed consists of three main steps. (i) we conduct a preliminary screening to reduce the number of candidate mediator genes by combining penalized regression in the outcome model with marginal modeling for each gene. Genes supported by both exposure-mediator and mediator-outcome associations are retained for further analysis. The marginal screening here is similar to the sure independence screening (SIS) approach \citep{SIS}. (ii) the selected genes are incorporated into the final mediation models to estimate IIEs. The outcome is modeled using linear regression, while mediator models are specified within a quasi-regression framework. This allows estimation of indirect effects corresponding to changes in average expression and changes in the proportion of cells expressing the gene. (iii) statistical significance is assessed using the joint significance (JS) test \citep{JS}, with false discovery rate (FDR) control to account for multiple testing across genes. We evaluate QuasiMed using real-data-driven simulations calibrated to observed single-cell characteristics and apply it to scRNA-seq data from the Religious Orders Study and Rush Memory and Aging Project (ROSMAP) to investigate mediation pathways associated with neurodegenerative processes. The paper concludes with a discussion of limitations and potential directions for future methodological development.

\section{Methods}\label{quasi:methods}
Subject index is suppressed in notation for simplicity throughout the entire section.
\subsection{Notation and Assumptions}
Let $Y$ be a continuous outcome and let $X$ denote the exposure. Let $\bm{Z} = \{Z_{1}, \dots  Z_{K} \}$ represent baseline covariates. For each cell $c = 1, \dots, C$ and each gene $g = 1, \dots, G$, let $M_{cg}$ denote the expression level. In this study, the interventional indirect effects (IIEs) are defined through contrasts of expected counterfactual outcomes \citep{Rubin1, counterfactual1} under exposure-induced changes in the mediator distribution \citep{counterfactual2, counterfactual3, IIEref}. Identification of the IIEs relies on the assumptions of no unmeasured confounding for the exposure-outcome, exposure-mediator, and mediator-outcome relationships. In contrast to natural indirect effects (NIE), the IIE has less assumptions and does not require cross-world independence assumptions between potential mediator and outcome values, and thus it can allow the mediators (i.e., genes) to be correlated whereas NIE cannot \citep{IIEref}.

\subsection{Model Specification}\label{quasi:modelspec}
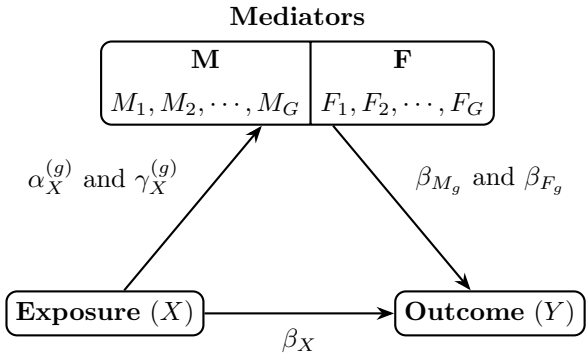
\begin{figure}[htb!]
\centering
\caption{A causal diagram for the mediation analysis with two co-mediators, $M_{g}$ and $F_{g}$. The interventional indirect effects quantify the mediation pathways through $M_{g}$ and $F_{g}$, and $\beta_{X}$ denotes the direct effect of exposure $X$ on outcome $Y$ that is not mediated by either mediator.}
\begin{tikzpicture}[node distance=2.5cm, thick, >=Stealth]
    \node (exposure) [draw, rectangle, rounded corners, align=center] {\textbf{Exposure} ($X$)};
    \node (outcome) [draw, rectangle, rounded corners, right=of exposure, align=center] {\textbf{Outcome} ($Y$)};
    \node (mediator) [
      draw,
      rectangle split,
      rectangle split horizontal,
      rectangle split parts=2,
      rounded corners,
      align=center,
      above=2.5cm of $(exposure)!0.5!(outcome)$,
      label={[font=\bfseries, yshift=2pt]above:Mediators}
    ]{
      \textbf{M}\\[4pt]
      $M_1, M_2, \cdots, M_G$
      \nodepart{second}
      \textbf{F}\\[4pt]
      $F_1, F_2, \cdots, F_G$
    };

    % Draw arrows and label the coefficients
    \draw[->] (exposure) -- (outcome) node[midway, below=2pt] {$\beta_{X}$}; % Slightly lowered
    \draw[->] (exposure) -- (mediator) node[pos=0.5, above left=2pt] {$\alpha_{X}^{(g)}$ and $\gamma_{X}^{(g)}$}; % Minimal offset
    \draw[->] (mediator) -- (outcome) node[pos=0.5, above right=2pt] {$\beta_{M_{g}}$ and $\beta_{F_{g}}$}; % Minimal offset
\end{tikzpicture}
\label{quasi:Fig1}
\end{figure}

We propose models corresponding to the causal diagram in Figure \ref{quasi:Fig1}. For each gene $g = 1, \dots , G$, we define two subject-level mediators. Let $\mathcal{A}_{g} = \{ c : M_{cg} > 0 \}$, $M_{g}= \frac{1}{|\mathcal{A}_{g}|} \sum_{c \in \mathcal{A}_{g}} M_{cg}$ denote the average expression across cells with non-zero expression, and let $F_{g}= \frac{1}{C} \sum_{c=1}^{C} I(M_{cg} > 0)$ denote the proportion of non-zero expression across cells. Let $X$ be the exposure and $\bm{Z}$ covariates.

The outcome model (Equation \ref{quasi:outcomemodel}) is 
\begin{align}
 Y = \beta_{0} + \sum_{g=1}^{G}\beta_{M_{g}} \log(M_{g})  & +  \sum_{g=1}^{G}\beta_{F_{g}} \text{logit}(F_{g})  +  \beta_{X}X + \bm{\beta_{Z}}^\top \bm{Z} + \varepsilon, \label{quasi:outcomemodel}
\end{align}
where $\beta_{M_{g}}$ and $\beta_{F_{g}}$ represent the effects of the expression level and expressed proportion (or equivalently zero proportion) of gene $g$ on the outcome, respectively, $\beta_{X}$ denotes the direct effect of $X$ on $Y$, $\bm{\beta_{Z}}^\top = (\beta_{Z_{1}}, \dots, \beta_{Z_{K}})$ represents the vector of coefficients for the covariates, and $\varepsilon$ is a normal random error term. 

For each gene $g$, we model the mean structures of $F_{g}$ and $M_{g}$ using a quasi-regression framework (Equations \ref{quasi:mediationmodel1} and \ref{quasi:mediationmodel2}) as follows
\begin{align}
\mathrm{E}\Bigl[\log(M_{g})\Bigr] &= \gamma_{0} + \gamma_{X}^{(g)}X + \bm{\gamma_{Z}}^{(g)\top} \bm{Z}.
\label{quasi:mediationmodel1}\\
\mathrm{E}\Bigl[\text{logit}(F_{g})\Bigr] &= \alpha_{0}+ \alpha_{X}^{(g)}X + \bm{\alpha_{Z}}^{(g)\top} \bm{Z}, \label{quasi:mediationmodel2}
\end{align}

The parameters $\alpha_{X}^{(g)}$ and $\gamma_{X}^{(g)}$ describe the effects of $X$ on the expressed proportion and expression level of gene $g$, respectively, and $\bm{\alpha_{Z}}^{(g)\top} = (\alpha_{Z_{1}}^{(g)}, \dots, \alpha_{Z_{K}}^{(g)})$ and $\bm{\gamma_{Z}}^{(g)\top} = (\gamma_{Z_{1}}^{(g)}, \dots, \gamma_{Z_{K}}^{(g)})$ are the corresponding covariate effects. This quasi-regression approach specifies only the mean relationships and avoids parametric assumptions on the mediator distributions. In scRNA-seq data, expressed proportions and expression levels are often highly variable \citep{zeroinf1, zeroinf2}, and standard parametric models may be restrictive.

Under this model, $M_{g}$ and $F_{g}$ can be written as $M_{g}(X, \bm{Z})$ and $F_{g}(X, \bm{Z})$ since they are functions of X and $\bm{Z}$. We can also write $Y$ as a function of all the variables: $Y(X, \bm{Z}, M_{1}(X, \bm{Z}), \dots, M_{G}(X, \bm{Z}), F_{1}(X, \bm{Z}), \dots, F_{G}(X, \bm{Z}))$. For simplicity, we can re-write it as $Y(X, M_{1}(X), \dots, M_{G}(X), F_{1}(X), \dots, F_{G}(X))$ where $z$ is dropped since it will be controlled in all analyses. 

\subsection{High-Dimensional Mediator Screening (Step I) }\label{quasi:screening}
Given the high dimensionality of the mediators, we implement a screening step prior to fitting the final mediation models. This step reduces the number of candidate genes while retaining those supported by both the outcome and mediator models. We first fit the outcome model in Equation \ref{quasi:outcomemodel} using Lasso regression, treating all aggregated mediators $M_g$ and $F_g$ as candidate predictors along with $X$ and $\bm{Z}$. Let $\mathcal{G}_Y$ denote the set of genes whose corresponding $M_g$ or $F_g$ terms are selected by the penalized outcome model. Next, for each gene $g$, we fit marginal linear models based on the mean structures specified in Equations \ref{quasi:mediationmodel1} and \ref{quasi:mediationmodel2}. Specifically, we regress $\text{logit}(F_g)$ and $\log(M_g)$ on $X$ and $\bm{Z}$ using quasi-regression. This marginal screening step is similar to the SIS approach \citep{SIS}, where variables are selected based on their marginal association with the exposure. Let $\mathcal{G}_F$ and $\mathcal{G}_M$ denote the sets of genes for which the standardized coefficients of $X$ are among the top $n/log(n)$ in the corresponding marginal models.

The final set of candidate genes is defined as 
\begin{align*}
    \mathcal{S} = (\mathcal{G}_{Y} \cap \mathcal{G}_{M}) \cup (\mathcal{G}_{Y} \cap \mathcal{G}_{F}). \nonumber
\end{align*}
For each $g \in \mathcal{S}$, the mediator component $M_{g}$ is included in the final outcome model if $g \in \mathcal{G}_{M}$, and $F_{g}$ is included if $g \in \mathcal{G}_{F}$. The final model therefore includes $X, \bm{Z}$, and the selected mediator terms.

\subsection{IIE Estimation (Step II)}\label{quasi:IIEestimation}

Based on Section \ref{quasi:modelspec}, the direct effect of $X$ on $Y$ (i.e., $X \rightarrow Y$) is estimated by the coefficient $\beta_{X}$ in the outcome model. We define two IIEs: (1) $\text{IIE}^{M_g}$, the indirect effect via $M_{g}$ (i.e., $X \rightarrow M_{g} \rightarrow Y$), and (2) $\text{IIE}^{F_g}$, the indirect effect via $F_{g}$ (i.e., $X \rightarrow F_{g} \rightarrow Y$). The IIE for $M_{g}$ when $X$ changes from $x_{1}$ to $x_{2}$ is:
\begin{align*}
    \text{IIE}^{M_g} = \E_{Z} \biggl[ Y\Bigl(x_{2}, W_{M_{g}}(x_{2}), W_{-M_{g}}(x_{1})\Bigr) - Y\Bigl(x_{2}, W_{M_{g}}(x_{1}), W_{-M_{g}}(x_{1})\Bigr) \biggr],
\end{align*}
where $W_{M_{g}}(x_{1})$ and $W_{M_{g}}(x_{2})$ denote a random draw from the distribution of $M_{g}(x_{1})$ and $M_{g}(x_{2})$, respectively, $W_{-M_{g}}(x_{1})$ denotes a random draw from the multivariate distribution of $\Bigl( M_{1}(x_{1}),\allowbreak F_{1}(x_{1}),\allowbreak M_{2}(x_{1}),\allowbreak F_{2}(x_{1}),\allowbreak \cdots,\allowbreak M_{g-1}(x_{1}),\allowbreak F_{g-1}(x_{1}),\allowbreak F_{g}(x_{1}),\allowbreak M_{g+1}(x_{1}),\allowbreak F_{g+1}(x_{1}),\allowbreak \cdots,\allowbreak M_{G}(x_{1}),\allowbreak F_{G}(x_{1}) \Bigr)$, and $ Y\Bigl(x_{2}, W_{M_{g}}(x_{2}), W_{-M_{g}}(x_{1})\Bigr)$ and $Y\Bigl(x_{2}, W_{M_{g}}(x_{1}), W_{-M_{g}}(x_{1})\Bigr)$ are counterfactual outcomes, and the subscript $Z$ means conditional on $Z$. Thus, we have
\begin{align*}
    \text{IIE}^{M_g} &= \beta_{X}x_{2} + \beta_{Z}^{\top}Z + \beta_{M_g}\E_{Z} \Bigl(\log \Bigl(W_{M_{g}}(x_{2})\Bigr) \Bigr) + \sum_{i \neq g}^{G} \beta_{M_i}\E_{Z} \Bigl(\log \Bigl(W_{M_{i}}(x_{1})\Bigr) \Bigr)  \\
    & \qquad+ \sum_{g = 1}^{G} \beta_{F_g}\E_{Z} \Bigl( \text{logit} \Bigl( W_{F_{g}}(x_{1}) \Bigr) \Bigr)  - \beta_{X}x_{2} - \beta_{Z}^{\top}Z - \beta_{M_g}\E_{Z} \Bigl( \log \Bigl( W_{M_{g}}(x_{1})\Bigr) \Bigr)  \\
    & \qquad - \sum_{i \neq g}^{G} \beta_{M_i}\E_{Z} \Bigl( \log \Bigl( W_{M_{i}}(x_{1})\Bigr) \Bigr)  - \sum_{g = 1}^{G} \beta_{F_g}\E_{Z} \Bigl( \text{logit} \Bigl( W_{F_{g}}(x_{1})\Bigr) \Bigr) \\
    &= \beta_{M_g}\E_{Z} \Bigl(\log \Bigl( W_{M_{g}}(x_{2})\Bigr) \Bigr) - \beta_{M_g}\E_{Z} \Bigl( \log \Bigl( W_{M_{g}}(x_{1})\Bigr) \Bigr) \\
    &= \beta_{M_g} \gamma^{(g)}_{X}.
\end{align*}
Similarly, the IIE for $F_{g}$ when $x$ changes from $x_{1}$ to $x_{2}$ is:
\begin{align*}
    \text{IIE}^{F_g} = \E_{Z} \biggl[ Y\Bigl(x_{2}, W_{F_{g}}(x_{2}), W_{-F_{g}}(x_{1})\Bigr) - Y\Bigl(x_{2}, W_{F_{g}}(x_{1}), W_{-F_{g}}(x_{1})\Bigr) \biggr],
\end{align*}
where $W_{F_{g}}(x_{1})$ and $W_{F_{g}}(x_{2})$ denote a random draw from the distribution of $F_{g}(x_{1})$ and $F_{g}(x_{2})$, respectively, and $W_{-F_{g}}(x_{1})$ denotes a random draw from the distribution of the vector $\Bigl( M_{1}(x_{1}),\allowbreak F_{1}(x_{1}),\allowbreak M_{2}(x_{1}),\allowbreak F_{2}(x_{1}),\allowbreak \cdots,\allowbreak M_{g-1}(x_{1}), \allowbreak F_{g-1}(x_{1}),\allowbreak M_{g}(x_{1}),\allowbreak M_{g+1}(x_{1}),\allowbreak F_{g+1}(x_{1}),\allowbreak \cdots,\allowbreak M_{G}(x_{1}),\allowbreak F_{G}(x_{1}) \Bigr)$. Thus, we have
\begin{align*}
    \text{IIE}^{F_g} &= \beta_{X}x_{2} + \beta_{Z}^{\top}Z +  \sum_{g = 1}^{G} \beta_{M_g}\E_{Z} \Bigl( \log \Bigl( W_{M_{g}}(x_{1})\Bigr) \Bigr) + \beta_{F_g}\E_{Z} \Bigl( \text{logit} \Bigl( W_{F_{g}}(x_{2})\Bigr) \Bigr) \\
    & \quad + \sum_{i \neq g}^{G} \beta_{F_g}\E_{Z} \Bigl( \text{logit} \Bigl( W_{F_g}(x_{1})\Bigr) \Bigr)  - \beta_{X}x_{2} - \beta_{Z}^{\top}Z - \sum_{g = 1}^{G} \beta_{M_g}\E_{Z} \Bigl( \log \Bigl( W_{M_{g}}(x_{1})\Bigr) \Bigr)  \\
    & \qquad - \beta_{F_g}\E_{Z} \Bigl( \text{logit} \Bigl( W_{F_{g}}(x_{1})\Bigr) \Bigr) - \sum_{i \neq g}^{G} \beta_{F_g}\E_{Z} \Bigl( \text{logit} \Bigl( W_{F_{g}}(x_{1})\Bigr) \Bigr)  \\
    &= \beta_{F_g} \Bigl( \E_{Z} \Bigl( \text{logit} \Bigl(W_{F_{g}}(x_{2})\Bigr) \Bigr) - \E_{Z} \Bigl( \text{logit} \Bigl(W_{F_{g}}(x_{1})\Bigr) \Bigr) \Bigr)  \\
    &= \beta_{F_g} \alpha^{(g)}_{X},
\end{align*}
where $\expit(\cdot)$ is the inverse of $\logit(\cdot)$ function.

\subsection{Hypothesis Testing (Step III)}\label{quasi:hypothesis}
To formally test these mediation effects, we establish the following hypothesis testing framework for each gene $g$:
\begin{align}
H_{0}^{M}: \beta_{M_{g}}\gamma_{X}^{(g)} &= 0, \nonumber \\
H_{0}^{F}: \beta_{F_{g}}\alpha_{X}^{(g)} &=0 \nonumber.
\end{align}
Under $H_{0}^{M}$, either $\beta_{M_{g}}$ or $\gamma_{X}^{(g)}$ is zero, indicating no mediation through $M_{g}$. Similarly, under $H_{0}^{F}$, either $\beta_{F_{g}}$ or $\alpha_{X}^{(g)}$ is zero, indicating no mediation through $F_{g}$. The joint significance (JS) test \citep{JS} is applied to evaluate whether a gene mediates the causal effect of $X$ on $Y$ through its expression level or the proportion of expressing cells. The JS test has been shown to control the type I error rate while maintaining statistical power in various omics studies \citep{JS.study1, JS.study2}. To test $H_{0}^{M}$ and $H_{0}^{F}$, we use the maximum of two p-values from the relevant coefficient estimates. Specifically, we define
\begin{align*}
    P_{\max_{g}}^{M} &= \max(P_{\beta_{M_{g}}}, P_{\gamma_{X}^{(g)}}), \nonumber \\
    P_{\max_{g}}^{F} &= \max(P_{\beta_{F_{g}}}, P_{\alpha_{X}^{(g)}}),
\end{align*}
where $P_{\beta_{M_{g}}}$ and $P_{\beta_{F_{g}}}$ are p-values obtained from the outcome model (Equations \ref{quasi:outcomemodel}), and $P_{\gamma_{X}^{(g)}}$ and $P_{\alpha_{X}^{(g)}}$ are p-values obtained from the mediator models (Equations \ref{quasi:mediationmodel1} and \ref{quasi:mediationmodel2}, respectively). Each p-value is calculated using the following test statistic:
\begin{align*}
    P_{\beta_{M_{g}}} &= 2\biggl\{1 - \Phi\biggl(\frac{|\hat{\beta}_{M_{g}}|}{\hat{\sigma}_{\beta_{M_{g}}}}\biggr) \biggr\}, \nonumber \\
    P_{\beta_{F_{g}}} &= 2\biggl\{1 - \Phi\biggl(\frac{|\hat{\beta}_{F_{g}}|}{\hat{\sigma}_{\beta_{F_{g}}}}\biggr) \biggr\}, \nonumber \\
    P_{\gamma_{X}^{(g)}} &= 2\biggl\{1 - \Phi\biggl(\frac{|\hat{\gamma}_{X}^{(g)}|}{\hat{\sigma}_{\gamma_{X}^{(g)}}}\biggr) \biggr\}, \nonumber \\
    P_{\alpha_{X}^{(g)}} &= 2\biggl\{1 - \Phi\biggl(\frac{|\hat{\alpha}_{X}^{(g)}|}{\hat{\sigma}_{\alpha_{X}^{(g)}}}\biggr) \biggr\}, \nonumber
\end{align*}
where $\hat{\beta}_{M_{g}}$, $\hat{\beta}_{F_{g}}$, $\hat{\gamma}_{X}^{(g)}$, and $\hat{\alpha}_{X}^{(g)}$ are the estimated regression coefficients from the fitted outcome and mediation models based on Equations \ref{quasi:outcomemodel} to \ref{quasi:mediationmodel2}, and $\hat{\sigma}_{\beta_{M_{g}}}$, $\hat{\sigma}_{\beta_{F_{g}}}$, $\hat{\sigma}_{\gamma_{X}^{(g)}}$, and $\hat{\sigma}_{\alpha_{X}^{(g)}}$ are their corresponding standard error estimates. The function $\Phi(\cdot)$ denotes the cumulative distribution function of the standard normal distribution. Given the large number of hypothesis tests across genes, it is important to control the false discovery rate (FDR). In this study, we apply the BH-adjusted p-value \citep{BH} to the p-values of the selected mediators.

\subsection{Boundary Value Handling for F-Mediation Model}\label{quasi:zerohandling}
In the proposed mediation model, $F_g$ denotes the proportion of positive counts for gene $g$ within a subject and therefore lies within a closed interval $[0, 1]$. When $F_{g}$ takes boundary values, it cannot be used directly in the model for $F_g$ (Equation \ref{quasi:mediationmodel2}). If a gene is expressed in all cells for some subjects, then $F_g = 1$. In this case, we set 
    \begin{align*}
        F_{g} = \min(F_{g}, 0.999).
    \end{align*}
Similarly, if a gene is not expressed at all in all cells for some subjects so that $F_{g} = 0$, we truncate 
    \begin{align*}
        F_{g} = \max(F_{g}, 0.001).
    \end{align*}
Finally, for genes with $F_{g}=1$ for all subjects, there is no variation in the zero proportion, and $F_{g}$ term is omitted from both mediation and outcome models.

\subsection{Performances Evaluation}
Performance was evaluated using statistical power and FDR, averaged over simulation replicates. In each replicate, the interventional-specific JS test was applied to every gene, separately assessing mediation through $M_{g}$ and $F_{g}$. Significance was determined at the $5\%$ level after BH correction. Power and FDR were computed as 
\begin{align*}
    \text{Power} &= \frac{TP}{T}, \qquad \text{FDR} = \frac{FP}{R}\mathbbm{1}{\{R > 0\}}, \nonumber 
\end{align*}
where $TP$ and $FP$ denote the numbers of true and false positives, $T$ is the total number of true mediators, and $R$ is the total number of discoveries.

\section{Simulation Study}\label{quasi:results}
Subject is indexed by $i$ in all notations throughout this section.

\subsection{Simulation Design}\label{quasi:sim.design}
Our simulation settings were guided by the ROSMAP single-cell data (see Section \ref{quasi:realdata} for details). We first summarized key design features from the preprocessed dataset, including the average number of cells per subject, number of genes measured per cell, and sample sizes. Data were then generated under the realistically comparable configurations while preserving distributional characteristics commonly observed in scRNA-seq, most notably the high frequency of zero counts. We considered the number of genes $g \in \{ 10000, 12000, 14000 \}$, sample sizes $n \in \{ 200, 300, 400\}$, and fixed the average number of cells per subject at $c = 40$.  The binary exposure variable $X$ was generated from a Bernoulli(0.5) distribution. For each subject $i$ and gene $g$, cell-level expression counts $M_{cg}^{(i)}$ were generated from a zero-inflated negative binomial (ZINB) distribution with gene-specific dispersion $\delta_{g}$, subject-specific mean $\mu_{ig}$, and subject-specific zero-inflation probability $\pi_{ig}$. The ZINB model was adopted to reflect the excess zeros commonly observed in scRNA-seq data. We linked $\mu_{ig}$ and $\pi_{ig}$ to the exposure through
\begin{align}
    \log(\mu_{ig}) &= \gamma_{X}^{(g)}X_{i}, \nonumber \\
    \text{logit}(\pi_{ig}) &= \alpha_{X}^{(g)}X_{i}, \nonumber
\end{align}
so that $\mu_{ig}$ and $\pi_{ig}$ represent subject-level parameters controlling the expected expression level and expressed proportion for gene $g$, respectively. $\alpha_{X}^{(g)}$ and $\gamma_{X}^{(g)}$ were generated as follows: $( \alpha_{X}^{(1)}, \dots, \alpha_{X}^{(4)}, \alpha_{X}^{(9)} \dots, \alpha_{X}^{(12)} )$ = $( \mathrm{Unif}(0.2,0.4), \mathrm{Unif}(0.2,0.4 ), \dots, \mathrm{Unif}( 0.2,0.4) )$ and $( \gamma_{X}^{(1)}, \gamma_{X}^{(2)}, \dots, \gamma_{X}^{(8)} )$ = $( \mathrm{Unif}( 0.2,0.4), \mathrm{Unif}(0.2,0.4 ), \dots, \mathrm{Unif}(0.2,0.4 ) )$, where $\mathrm{Unif}(\cdot,\cdot)$ denote uniform distribution. The rest of $\alpha_{X}^{(g)}$ and $\gamma_{X}^{(g)}$ are set to be zero.

Based on the model specification above, for each cell, a zero count was first generated from a Bernoulli distribution with probability $\pi_{ig}$. Otherwise, counts were generated from a negative binomial distribution with mean $\mu_{ig}$ and dispersion parameter $\delta_{g}$, where $\delta_{g} \sim \mathrm{Unif}(0.6, 1.2)$. This defines
\begin{align*}
    M_{cg}^{(i)} \sim ZINB(\mu_{ig}, \delta_{g}, \pi_{ig}).
\end{align*}
Subject-level summaries were obtained by aggregating across cells as follows
\begin{align*}
    M_{g}^{(i)} &= \frac{1}{|\mathcal{A}_{g}|} \sum_{c \in \mathcal{A}_{g}^{i}} M_{cg}^{(i)}, \qquad
    F_{g}^{(i)} = \frac{1}{C} \sum_{c = 1}^{C} I\Bigl(M_{cg}^{(i)} > 0 \Bigr),
\end{align*}
where $M_{g}^{(i)}$ denotes the average expression and $F_{g}^{(i)}$ denotes the proportion of non-zero counts for gene $g$ in subject $i$.

After generating the cell-level counts, we applied the preprocessing steps described in Section \ref{quasi:zerohandling}. Genes that are not expressed at all across all subjects were removed. To keep the expressed proportions away from the boundaries, we truncated
\begin{align*}
    \hat{F}_{g}^{(i)} = \min(\max(F_{g}^{(i)}, 0.001), 0.999).    
\end{align*}
This prevents numerical instability in modeling $F_{g}^{(i)}$. 

For generating $Y$, we used the following equation: $Y=\beta_{0} + \sum_{g=1}^{G} \beta_{M_{g}}\log(M_{g}) + \sum_{g=1}^{G} \beta_{F_{g}}\text{logit}(F_{g}) + \beta_{X}X+\epsilon$, where $(\beta_{M_{1}}, \beta_{M_{2}}, \dots, \beta_{M_{8}} ) = ( \mathrm{Unif}(0.8,1.1 ), \mathrm{Unif}( 0.8,1.1), \dots, \mathrm{Unif}(0.8,1.1 ) )$ and $(\beta_{F_{1}}, \dots, \beta_{F_{4}},\beta_{F_{9}} \dots, \beta_{F_{12}} ) = ( \mathrm{Unif}(0.8,1.1), \mathrm{Unif}(0.8,1.1 ), \dots, \mathrm{Unif}(0.8,1.1) )$ and the rest of $\beta_{M_{g}}$ and $\beta_{F_{g}}$ were set to be zero. In addition, $\beta_{X}=3$ and the random error $\epsilon$ was generated from the standard normal distribution. Notice that under this parameter setting, the true mediators are $M_{1}$ to $M_{8}$, $F_{1}$ to $F_{4}$, and $F_{9}$ to $F_{12}$.

To benchmark the performance of QuasiMed, we also considered a naïve approach based on marginal modeling. Under this approach, each gene was analyzed separately. For the marginal outcome model, the outcome was regressed on the gene and $X$. For the marginal mediation model, each gene was regressed on $X$. No preliminary screening was performed. The same JS test with BH adjustment was applied. Simulation results were averaged over 100 replicates.

\subsection{Simulation Results}\label{quasi:sim.results}

Table \ref{quasi:table1} summarizes simulation results under real-data-driven high-dimensional settings with sample sizes $n \in \{200,300,400\}$, gene sizes $g \in \{10000,12000,14000\}$, and cell size $c=40$. Power(M) and FDR(M) correspond to testing $\mathrm{IIE}^{M_g}$, and Power(F) and FDR(F) correspond to testing $\mathrm{IIE}^{F_g}$. For $n=200$, QuasiMed achieves high power for both $M$ and $F$ models across all gene sizes, with Power(M) ranging from 0.888 to 0.988 and Power(F) ranging from 0.910 to 0.995. FDR is well controlled in all three settings, remaining below 0.03 for both $M$ and $F$ models. When the sample size increases to $n=300$, QuasiMed continues to maintain high power and stable FDR control. Across gene sizes, Power(M) ranges from 0.915 to 0.988 and Power(F) ranges from 0.910 to 0.989, while both FDR(M) and FDR(F) stay below the nominal $5\%$ level. For $n=400$, QuasiMed achieves high power in both models, with Power(M) between 0.982 and 1.000 and Power(F) between 0.964 and 0.999, while still maintaining stable FDR control across all gene sizes. It is worth noting that the data were not generated exactly under the proposed models, which demonstrates the robustness of our approach.

The naïve approach, in contrast, controls FDR across all scenarios considered but remains highly conservative, with FDR equal to 0 in every setting. Its power is consistently and substantially lower than that of QuasiMed for both $M$ and $F$ models. For example, when $n=200$, the naïve method attains Power(M) between 0.327 and 0.508 and Power(F) between 0.275 and 0.608, and at $n=400$ its power remains far below that of QuasiMed, with Power(M) between 0.258 and 0.386 and Power(F) between 0.311 and 0.433. Overall, these results are consistent with the conservative testing behavior of the naïve procedure. 

For computation time, QuasiMed consistently outperforms the naïve approach in every scenario considered. The computational advantage is especially noticeable as the gene size increases, reflecting the benefit of the screening step, which combines penalization in the outcome model and marginal modeling for the $M$ and $F$ models, thereby reducing the number of tested hypotheses relative to the naïve method. All simulations were run on the University of Florida high-performance Linux cluster, HiPerGator 3.0, using 15 CPU cores and 3 GB of RAM per node.

\begin{table*}[!htb]
\centering
\small
\setlength{\tabcolsep}{3pt}
\caption{Simulation results comparing QuasiMed and the naïve approach under real-data-like parameter settings. }
\label{quasi:table1}
\begin{tabular}{@{} cc cccccc @{}}
\toprule
 $n$ & $G$ & Methods & Power(M) & Power(F) & FDR(M) & FDR(F) & \makecell{Average \\ Comp. \\ Time \\ (in secs)} \\
\midrule

%\multicolumn{9}{@{}l}{Scenario IV} \\

 200 & 10000 & QuasiMed & 0.988 & 0.995 & 0.01 & 0.022 & 81.027   \\
 &    &  Naïve & 0.508 & 0.608 & 0.000 & 0.000 & 148.215 \\
       & 12000 & QuasiMed & 0.906 & 0.984 & 0.0203 & 0.0042 & 186.673 \\
        &  & Naïve & 0.327 & 0.518 & 0.000 & 0.000 & 337.718 \\
   &  14000 & QuasiMed & 0.888 & 0.91 & 0.03 & 0.022 & 140.117 \\
   &    & Naïve & 0.39 & 0.275 & 0.000 & 0.000 & 260.046 \\
\midrule
 300 & 10000 & QuasiMed & 0.988 & 0.989 & 0.02 & 0.024 & 124.177   \\
 &    &  Naïve & 0.234 & 0.321 & 0.000 & 0.000 & 217.063 \\
       & 12000 & QuasiMed & 0.915 & 0.91 & 0.016 & 0.011 & 108.305 \\
        &  & Naïve & 0.116 & 0.212 & 0.000 & 0.000 & 193.992 \\
   &  14000 & QuasiMed & 0.93 & 0.985 & 0.021 & 0.024 & 136.505 \\
   &    & Naïve & 0.153 & 0.316 & 0.000 & 0.000 & 249.067 \\
\midrule
 400 & 10000 & QuasiMed &  1.000 & 0.964 & 0.023 & 0.026 & 67.069   \\
 &    &  Naïve & 0.386 & 0.311 & 0.000 & 0.000 & 119.789 \\
       & 12000 & QuasiMed & 0.993 & 0.999 & 0.021 & 0.017 & 176.37 \\
        &  & Naïve & 0.258 & 0.433 & 0.000 & 0.000 & 308.268 \\
   &  14000 & QuasiMed & 0.982 & 0.996 & 0.013 & 0.021 & 251.338 \\
   &    & Naïve & 0.308 & 0.383 & 0.000 & 0.000 & 444.343 \\
\bottomrule
\end{tabular}

\begin{tablenotes}
\item [] Abbreviations: Power(M), power for M models; Power(F), power for F models; FDR(M), false discovery rate for M models; FDR(F), false discovery rate for F models.
\end{tablenotes}
\end{table*}

\section{Real Data Application}
\subsection{ROSMAP Study and Data Description}\label{quasi:realdata}
The Religious Orders Study and the Rush Memory and Aging Project (ROSMAP) provide postmortem brain tissue and clinical characterization for older adults and have been used to generate large single-nucleus transcriptomic datasets. Recent work from this cohort profiled nuclei from prefrontal cortex samples across 427 participants and evaluated relationships between cell-type specific molecular features, Alzheimer's disease (AD) pathology, and cognitive impairment \citep{ROSMAPoriginal}. These analyses reported vulnerable neuronal subtypes and broader shifts in non-neuronal populations, including glial and vascular-related compartments, along with gene expression programs linked to neuronal stress and survival.

We focus on vascular and epithelial cells because they are rare in these datasets but are repeatedly implicated in ROSMAP-based single-cell analyses of Alzheimer’s disease. One study using AD-affected brain regions from the ROSMAP cohort reported higher abundance of vascular and epithelial cells in late-stage disease, including in entorhinal cortex, hippocampus, and prefrontal cortex \citep{ROSMAP1}. In another ROSMAP study that profiled seven major cell types, vascular and epithelial cells accounted for approximately $0.8\%$ of all cells \citep{ROSMAPdatapaper}. Their low abundance and dispersion across anatomy make them difficult to study, and they have been noted as understudied in this context \citep{ROSMAP2}. Prior work has also described associations between cerebrovascular dysfunction, AD pathology, and early cognitive decline \citep{ROSMAP3}.

In this analysis, postmortem interval (PMI) is treated as the outcome. PMI is the time between the recorded time of death and the time of tissue collection and preservation through freezing or fixation. Although PMI is often treated as a technical variable, we treat it here as a biologically relevant outcome that may reflect processes occurring around the agonal period and immediately after death \citep{postmortem1, postmortem2}. Variability in PMI may relate to physiological stress, inflammation, agonal state, or pre-terminal pathology, and we consider the possibility that these factors are reflected in molecular and cellular features measured in postmortem tissue. We test whether pathologic AD status is associated with PMI and whether vascular or epithelial cell composition mediates this association, focusing on aggregated abundance and sparsity patterns. Sex is included as an additional covariate. In ROSMAP, the pathologic diagnosis and the single-cell RNA-seq measurements are obtained at or shortly after the time of death, which coincides with the start of PMI measurement \citep{ROSMAPdatapaper, ROSMAPtiming}. This alignment motivates treating PMI as the outcome in our mediation framework.

\begin{figure}[hbt!]
\centerline{\includegraphics[scale=0.14]{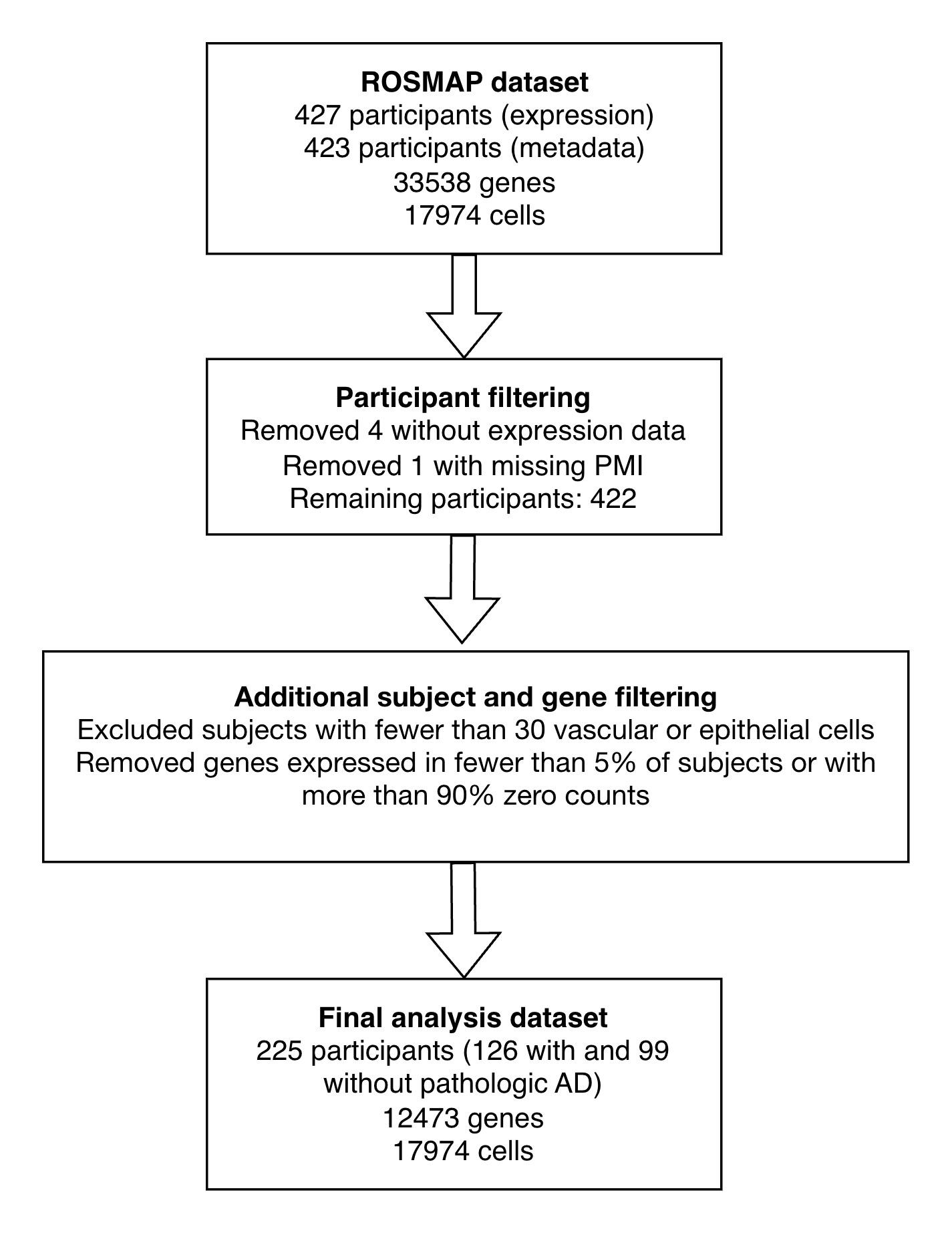}}
\caption{Flowchart of preprocessing and filtering of the ROSMAP data used in the QuasiMed analysis.}
\label{quasi:Fig2}
\end{figure}

We used pre-processed single-cell gene count data for vascular and epithelial cells (\texttt{Vasculature\_cells.rds}) together with de-identified clinical metadata (\texttt{individual\_metadata\_deidentified.tsv}) from the ROSMAP study. Data were obtained from the AD/Aging Brain Atlas (\url{https://compbio.mit.edu/ad_aging_brain/}
) and log-normalized using Seurat's \texttt{LogNormalize} procedure. The gene count matrix (33,538 genes by 17,974 cells) was loaded as a Seurat object in \texttt{Seurat}. The expression data included 423 participants, whereas the metadata listed 427. Four participants without corresponding expression data were removed, and one additional participant was excluded due to missing PMI, resulting in 422 participants. The ROSMAP reference paper did not report $M_g$ and $F_g$ directly but summarized cell-level expression using gene-level aggregates within cell types \citep{ROSMAPdatapaper}. We then applied additional filtering criteria. Subjects with fewer than 30 vascular or epithelial cells were excluded. Genes were required to be expressed in at least $5\%$ of subjects, and genes with more than $90\%$ zero counts were removed. The final dataset included 225 subjects ($n = 126$ with and $n = 99$ without a pathologic diagnosis of AD) and 12,473 genes. These preprocessing and filtering steps are summarized in the flowchart in Figure \ref{quasi:Fig2}.

\subsection{Analysis of the ROSMAP Dataset}
Table \ref{quasi:table3} summarizes the top-ranked genes from the $M$ model (i.e., $X \rightarrow M \rightarrow Y$) and the $F$ model (i.e., $X \rightarrow F \rightarrow Y$). No gene was shared across the two pathways, indicating that the mediating gene sets differed between the $M$ and $F$ models. Of the three $M$ mediators, FAM199X had a negative indirect effect ($p_{max}$ = 0.002), the estimated pathway effects of $X \rightarrow M$ and $M \rightarrow Y$ were 1.663 and -0.457, respectively. FAM199X has been discussed in prior studies in relation to neuronal hypermethylation \citep{FAM199X1} and circadian regulation \citep{FAM199X2}. However, based on the Harmonizome 3.0 database \citep{harmonizome}, a multi-omics data integration resource that curates gene and protein information across 138 published datasets, no study has directly characterized its biological function. Its role remains largely unexplored in AD/ADRD research. The $M$ model estimated a positive indirect effect for CD226 ($p_{max} = 0.054$), with estimated pathway effects along $X \rightarrow M$ and $M \rightarrow Y$ equal to $-0.860$ and $-0.429$, respectively. A recent transcriptomic study \citep{CD226} showed that a variant in CD226 is associated with multiple sclerosis, a chronic neurological autoimmune disease. For C2orf42 gene, which has been reported as a neuroimmune factor in AD patients \citep{C2orf40}, the estimated component effects along the $X \rightarrow M$ and $M \rightarrow Y$ paths were $-0.774$ and $-0.402$, respectively, resulting in a positive indirect effect ($p_{max} = 0.059$).

Of the $F$ mediators, GDPD1 had a positive indirect effect ($p_{max} = 0.015$) with estimated pathway effects of 1.139 for $X \rightarrow F$ and 0.468 for $F \rightarrow Y$, respectively. GDPD1 has been selected as an AD-associated biomarker in the analysis of peripheral blood markers for early-phase AD diagnosis \citep{GDPD1}. SLC4A8 had a negative indirect effect ($p_{max} = 0.019$) with pathway effects of -1.06 for $X \rightarrow F$ and 0.45 for $F \rightarrow Y$, respectively. A recent AD study \citep{SLC4A8} performed a transcriptome-wide association study using the joint-tissue imputation approach and a Mendelian randomization framework, where SLC4A8 was identified as a novel potential AD-related gene, although its function in AD has not yet been reported. CORO1B showed a negative indirect effect ($p_{max} = 0.051$), with pathway effects of 0.943 for $X \rightarrow F$ and -0.407 for $F \rightarrow Y$, respectively. ZNF337.AS1 gene showed a negative indirect effect ($p_{max} = 0.055$), with estimated pathway effects of -1.783 and 0.424. CORO1B was recently reported in a differential methylation study as one of the 30 hypomethylated CpG sites identified in peripheral blood leukocytes from AD patients compared to controls \citep{CORO1B}. ZNF337.AS1 was also reported in a differential methylation study in AD patients compared to controls \citep{ZNF337AS1}. While these genes have previously appeared in association-based studies, our results suggest that they may also play a mediating role in the pathway identified in our analysis. The ROSMAP data analysis took approximately 4 minutes and 3 seconds on the University of Florida high-performance Linux cluster, HiPerGator 3.0, using 15 CPU cores and 3 GB of RAM per node.

\begin{table*}[!htb]
\centering
\normalsize
\setlength{\tabcolsep}{11pt}
\caption{Top-ranked genes in $M$ and $F$ models (i.e., $\hat{\beta}_{M_{g}}\hat{\gamma}_{X}^{(g)} > 0$ and $\hat{\beta}_{F_{g}}\hat{\alpha}_{X}^{(g)} > 0$) identified from the analysis of ROSMAP data using the proposed method}
\label{quasi:table3}
\begin{tabular}{@{} c cc cc @{}}
\toprule
 M Model & Gene & $\hat{\beta}_{M_{g}}$ & $\hat{\gamma}_{X}^{(g)}$ & $P_{\max_{g}}^{M}$  \\
\midrule 
 & FAM199X & 1.663 & -0.457 & 0.002  \\
 & CD226 & -0.860 & -0.429 & 0.054  \\
 & C2orf42 & -0.774 & -0.402 & 0.059  \\
\midrule
 F Model & Gene & $\hat{\beta}_{F_{g}}$ & $\hat{\alpha}_{X}^{(g)}$ & $P_{\max_{g}}^{F}$  \\
 \midrule
 & GDPD1 & 1.139 & 0.468 & 0.015  \\
 & SLC4A8 & -1.06 & 0.45 & 0.019  \\
 & CORO1B & 0.943 & -0.407 & 0.051  \\
 & ZNF337.AS1  & -1.783 & 0.424 & 0.055   \\
\bottomrule
\end{tabular}
\begin{tablenotes}
\item [] Top hits with unadjusted p-values less than or close to 0.05 are reported in the table.
\end{tablenotes}
\end{table*}

\section{Discussion}
In this study, we examine indirect effects through two gene-level components, the average expression level ($\text{IIE}^{M_g}$) and the proportion of expressing cells ($\text{IIE}^{F_g}$), corresponding to a decomposition of the IIEs. To estimate these effects, we use a quasi-regression-based approach that does not require any distributional assumptions for the mediators. A closely related work to the proposed method is MedZIsc \citep{MedZIsc}, but the present study differs in several aspects. Firstly, while both approaches adopt similar link functions on the response variable in mediation models, MedZIsc specifies NB and beta distributions for modeling gene expression and expression proportions, whereas QuasiMed models the transformed quantities directly without requiring those distributional assumptions. Secondly, $M_{g}$ is defined differently. MedZIsc averages expression over all cells, while QuasiMed averages only over non-zero cells.

In our simulations that mimic the ROSMAP data structure with similar gene size, sample size, and cell counts per subject, QuasiMed showed high power, controlled FDR at the nominal level, and required less computation time. In the ROSMAP data analysis, QuasiMed identified genes whose average expression levels or expressed proportions were involved in the pathway between AD pathology and PMI. Several of these genes have been reported in recent studies as neuroimmune or AD-related markers. Among them, FAM199X from the M component has been noted in multi-omics study database \citep{harmonizome}, even though its role in AD/ADRD is not yet well characterized. These findings point to candidate genes for further study in the context of AD pathology.

While the proposed framework is useful, several extensions remain. In the current formulation, we focus on a continuous outcome. The quasi-regression structure for the mediators remains unchanged, but the outcome model can be adapted to other types of responses. For example, a logistic model may be used for binary outcomes, and Cox or accelerated failure time models may be considered for survival outcomes. A natural extension would be to systematically investigate these alternative outcome settings within the same mediation framework. Moreover, our proposed framework includes a screening step through Lasso penalization combined with marginal modeling. Further refinement of the penalization step may consider alternative choices such as minimax concave penalty \citep{MCP} or elastic net \citep{elnet}. Lastly, the BH adjustment was used for multiple testing after the JS test. Future work may consider alternative multiple testing procedures, such as the q-value method \citep{qvalue}.

%\section{Supplementary Information}\label{smap:supplementary}
%See the figure in Supplementary Materials document.

\backmatter

%\bmhead{Supplementary information}

%If your article has accompanying supplementary file/s please state so here. 

%Authors reporting data from electrophoretic gels and blots should supply the full unprocessed scans for key as part of their Supplementary information. This may be requested by the editorial team/s if it is missing.

%Please refer to Journal-level guidance for any specific requirements.

\subsection*{Acknowledgments}
S.A. would like to express sincere gratitude for the extensive support received from the Department of Otolaryngology - Head and Neck Surgery.

\section*{Declarations}
\subsection*{Author contributions}
Conceptualization: S.A., Z.L. Methodology: S.A., Z.L. Simulation: S.A., D.P., Z.L. Software: S.A., D.P., Z.L. Formal Analysis: S.A., D.P., Z.L. Writing - Original Draft: S.A. Review and Editing: D.P., L.C., M.V.G., P.R., Z.L.

\subsection*{Competing interests}
The authors declare that they have no competing interests.

\subsection*{Code availability}
The R package is freely available on the GitHub repository at \url{https:/github.com/sjahnn/QuasiMed}). Please reach out to the corresponding author (Seungjun Ahn, seungjun.ahn@mountsinai.org) if you have any further inquiries.

%\subsection*{Funding}

%\subsection*{Ethics approval} 
%Not applicable.
%\subsection*{Consent to participate}
%Not applicable.
%\subsection*{Consent for publication}
%Not applicable.

\noindent

\bibliography{sn-bibliography}% common bib file


\newpage

\newpage

% Turn off if Appendix not used below:
%\begin{appendices}

%An appendix contains supplementary information that is not an essential part of the text itself but which may be helpful in providing a more comprehensive understanding of the research problem or it is information that is too cumbersome to be included in the body of the paper.




%\end{appendices}

\newpage 

\end{document}